\documentclass[a4paper,11pt]{article}

\usepackage[labelfont=bf,labelsep=period]{caption}
\usepackage[nodayofweek]{datetime}
\usepackage{enumitem}
\usepackage{float}
\usepackage[margin=2.5cm]{geometry}
\usepackage{graphicx}
\usepackage{numbertabbing}
\usepackage{times}
\usepackage{url}
\usepackage{hyperref}
\usepackage{xspace}
\usepackage{amsmath} 
\allowdisplaybreaks[2]          
\usepackage{amssymb} 
\usepackage{amsthm} 
\newtheoremstyle{plain-boldhead}
  {\topsep}
  {\topsep}
  {\itshape}
  {}
  {\bfseries}
  {.}
  { }
  {\thmname{#1}\thmnumber{ #2}\thmnote{ (\bfseries #3)}}
\newtheoremstyle{definition-boldhead}
  {\topsep}
  {\topsep}
  {\normalfont}
  {}
  {\bfseries}
  {.}
  { }
  {\thmname{#1}\thmnumber{ #2}\thmnote{ (\bfseries #3)}}
\theoremstyle{plain-boldhead}
\newtheorem{theorem}{Theorem}

\newtheorem{lemma}[theorem]{Lemma}
\newtheorem{corollary}[theorem]{Corollary}

\theoremstyle{definition-boldhead}
\newtheorem{definition}{Definition}
\newtheorem{remark}{Remark}

\newdateformat{simple}{\THEDAY\ \monthname[\THEMONTH]\ \THEYEAR}
\simple

\floatstyle{ruled}
\newfloat{algo}{htbp}{algo}
\floatname{algo}{Algorithm}

\def \ifempty#1{\def\temp{#1} \ifx\temp\empty }

\newcommand{\E}{\mathrm{E}}

\newcommand{\var}[1]{\textit{#1}}
\newcommand{\op}[1]{\textsl{#1}}

\newcommand{\true}{\textsc{true}\xspace}

\newcommand{\CA}{\ensuremath{\mathcal{A}}\xspace}

\newcommand{\CD}{\ensuremath{\mathcal{D}}\xspace}
\newcommand{\CE}{\ensuremath{\mathcal{E}}\xspace}

\newcommand{\CP}{\ensuremath{\mathcal{P}}\xspace}

\newcommand\tx{\var{tx}\xspace}

\newcommand\bl{\var{b}\xspace}

\newcommand\babbcast{\op{bab-broadcast}}

\newcommand\babdel{\op{bab-deliver}}
\newcommand\babmine{\op{bab-mined}\xspace}

\newcommand\VT{\op{VT}\xspace}
\newcommand\VB{\op{VB}\xspace}

\newcommand\ntx{\ensuremath{m }\xspace}
\usepackage{todonotes}
\setuptodonotes{fancyline, inline, color=red!30}
\usepackage[commandnameprefix=ifneeded,commentmarkup=uwave]{changes}

\definechangesauthor[name=nacho, color=red]{nacho}

\begin{document}

\title{\bf We will DAG you}

\author{Ignacio Amores-Sesar\\
  University of Bern\\
  \url{ignacio.amores@unibe.ch}
  \and Christian Cachin\\
  University of Bern\\
  \url{christian.cachin@unibe.ch}
}

\date{\today}

\maketitle

\begin{abstract}\noindent
  DAG-based protocols have been proposed as potential solutions to the latency and throughput limitations of traditional permissionless consensus protocols. However, their adoption has been hindered by security concerns and a lack of a solid foundation to guarantee improvements in both throughput and latency. In this paper, we present a construction that rigorously demonstrates how DAG-based protocols can achieve superior throughput and latency compared to chain-based consensus protocols, all while maintaining the same level of security guarantees.

\end{abstract}

\section{Introduction}
\label{sec:intro}

In the ever-evolving landscape of distributed systems, achieving consensus among a set of processes has become a fundamental challenge that has garnered significant attention in recent years. Consensus protocols are a universal primitive in distributed computing, ensuring that a network of interconnected processes can collectively agree on a shared state despite potential failures or malicious actors. However, as the demands on distributed systems continue to grow, the need for consensus protocols that can deliver both higher throughput and lower latency has become increasingly pressing. This need is particularly relevant in permissionless consensus protocols as used by cryptocurrencies and blockchain protocols, which face stringent demands on their throughput and latency.

Traditional consensus protocols have exhibited considerable advancements in both throughput and latency since the first practical consensus protocols~\cite{DBLP:journals/tocs/Lamport98,DBLP:journals/tocs/CastroL02}. One of the most promising lines of work are DAG consensus protocols as introduced by the ``All you need is DAG'' paper~\cite{DBLP:conf/podc/KeidarKNS21} and subsequently extended by Narwhal and Tusk~\cite{DBLP:conf/eurosys/DanezisKSS22}, Bullshark~\cite{DBLP:conf/ccs/SpiegelmanGSK22}, and Cordial Miners~\cite{DBLP:conf/wdag/KeidarNPS23}. A common characteristic of these protocols is their capacity to enable every participant to generate blocks that reference previous blocks, forming a \emph{directed acyclic graph} (\emph{DAG}). In permissionless protocols like Bitcoin~\cite{nakamoto2019bitcoin}, every process (miner) can create a block upon successfully solving the cryptographic puzzle. Therefore, the concept of constructing a DAG that is later ordered, as proposed by Keidar et al.~\cite{DBLP:conf/podc/KeidarKNS21}, holds the potential to enhance the throughput and latency of permissionless consensus protocols. In essence, DAG protocols may surpass traditional permissionless consensus protocols, which form a chain.

The evident approach to improving the throughput of chain protocols is to increase the block ratio, i.e., the number of block produced per unit of time, effectively accelerating the execution of the protocol as there is less time between created blocks. This goal can be pursued by lowering the difficulty in \emph{Proof-of-Work} (\emph{PoW}) protocols. However, increasing the block ratio may harm the protocol since it elevates the likelihood of forks—situations where two different processes create blocks extending the chain. An abandoned block is one that is never output by the protocol, whenever a chain protocol forks, an abandoned block is produced. Therefore, despite the increased number of generated blocks, the number of abandoned blocks concurrently rises, adversely affecting the protocol's throughput. Moreover, it is imperative to recognize that the block ratio cannot be augmented arbitrarily without compromising the protocol's security.

In this paper, we introduce a construction that takes as input a DAG-based protocol or a chain protocol $\Pi$, which may produce abandoned blocks, and produces a new DAG protocol $\Pi'$ with the property that every created block is eventually output. Specifically, $\Pi'$ creates the same number of blocks as the base protocol $\Pi$ and outputs \emph{every} created block of $\Pi$. We show that the safety and liveness of $\Pi'$ reduces to the safety and liveness of $\Pi$. In simpler terms, $\Pi'$ is as safe and live as $\Pi$.
Furthermore, we establish that $\Pi'$ has lower or equal \emph{latency} as $\Pi$, while achieving strictly higher \emph{throughput}. Our main contribution lies in a formal proof that chain protocols cannot achieve optimal throughput, i.e., for any chain protocol~$\Pi$, there is a DAG protocol~$\Pi'$ that is safe and life under the same assumptions as $\Pi$, with the same or better latency and better throughput.

\section{Related work}
\label{sec:related}

DAG protocols represent a recent breakthrough within the domain of permissioned consensus protocols~\cite{DBLP:conf/podc/KeidarKNS21,DBLP:conf/eurosys/DanezisKSS22,DBLP:conf/ccs/SpiegelmanGSK22,DBLP:conf/wdag/KeidarNPS23}.
While DAG protocols have been previously introduced in the permissionless context, their adoption and success have been somewhat restrained due to their inherent complexity when compared to traditional chain protocols. Several well-known DAG protocols have exhibited vulnerabilities, highlighting challenges in their success. For instance, IOTA~\cite{DBLP:journals/candie/PopovSF19}, one of the pioneering DAG protocols, has been susceptible to vulnerabilities such as Parasite-chain attacks~\cite{DBLP:journals/candie/PopovSF19,DBLP:journals/corr/abs-2004-13409}. Another promising protocol, GhostDAG~\cite{DBLP:conf/aft/SompolinskyWZ21}, has also revealed vulnerabilities in its design~\cite{DBLP:conf/usenix/LiLZYWYXLY20}. Even Avalanche~\cite{DBLP:journals/corr/abs-1906-08936}, the most successful DAG protocol in terms of market capitalization, originally had vulnerabilities in its design~\cite{DBLP:conf/opodis/Amores-SesarCT22}.

An intriguing DAG protocol to note is Conflux\cite{DBLP:conf/usenix/LiLZYWYXLY20}, which leverages the GHOST consensus rule~\cite{DBLP:conf/fc/SompolinskyZ15} and augments blocks with additional references to transform a chain protocol into a DAG. Li et al.~\cite{DBLP:conf/usenix/LiLZYWYXLY20} have demonstrated that Conflux's security is directly inherited from the security of GHOST. However, it is worth mentioning that the GHOST protocol has exhibited lower resilience than other consensus protocols in the presence of network malfunctions~\cite{DBLP:conf/dsn/NatoliG17,DBLP:conf/ccs/BagariaKTFV19}.

Our contribution to this landscape is a formal proof of the superior performance of DAG protocols, facilitated by a construction that can be conceptualized as an extension of the Conflux construction~\cite{DBLP:conf/usenix/LiLZYWYXLY20}. Specifically, when we instantiate the throughput closure using GHOST~\cite{DBLP:conf/fc/SompolinskyZ15}, we arrive at Conflux~\cite{DBLP:conf/usenix/LiLZYWYXLY20}.

\section{Abstractions}
\label{sec:abstractions}
We consider a set of $n$
\emph{processes} $\CP =\{P_1, P_2, \ldots \}$ that interact with each other by exchanging
messages through the network. A protocol $\Pi$ for $\CP$ consists of a collection of programs
with instructions for all processes. In particular we are interested in the study of \emph{chain protocol}
and \emph{DAG protocol} protocols, i.e., protocol that rely on a chain or a DAG to deliver blocks. These two concepts are formally defined below.

Chain and DAG protocols are pivotal tools employed to establish robust and secure ledgers, and as such, they must adhere to specific fundamental requirements.

Traditionally, the gold standard concept is \emph{atomic broadcast}~\cite{DBLP:books/daglib/0025983}, which ensures that all processes
deliver the same set of transactions in the same order. In this paper, we consider a variant of this abstraction that includes the concept of a \emph{block} in the interface and properties~\cite{DBLP:journals/corr/abs-2307-02954}.
Processes broadcast transactions and deliver blocks using the events 
$\babbcast(\tx)$ and $\babdel(\bl)$, respectively,
where block \bl contains a sequence of transactions $[\tx_1, \ldots, \tx_{\ntx}]$.
The protocol outputs an additional event $\babmine(\bl, P)$, 
which signals that block~\bl has been \emph{mined} by process $P$, where $P$ is defined as the \emph{miner} of \bl. The event $\babmine(\bl, P)$ can be understood as the creation of block $\bl$ by party $P$.
Notice that $\babmine(\bl, P)$ signals only the creation of a block and not its delivery.
In addition to predicate $\VT()$ that determines the validity of a transaction, we also equip our protocol
with a validity predicate $\VB()$ to be applied to blocks.
These predicates and function are determined by the higher-level application or protocol.

\begin{definition}\label{def:bab}
  A protocol implements \emph{block-based atomic broadcast} with validity predicates $\VT()$ and $\VB()$ if it satisfies the following properties, except with negligible probability:
 \begin{description}
  \item[Validity:] If a correct process invokes a $\babbcast(\tx)$, then every correct process eventually outputs $\babdel(\bl)$, for some block \bl that contains $\var{tx}$.
  \item[No duplication:] No correct process outputs $\babdel(\bl)$ more than once.
  \item[Integrity:] If a correct process outputs $\babdel(\bl)$, then it has previously output $\babmine(\bl, \cdot)$ exactly once.
  \item[Agreement:] If some correct process outputs $\babdel(\bl)$, then eventually every correct process outputs $\babdel(\bl)$.
  \item[Total order:] Let $\bl$ and $\bl'$ be blocks, and $P_i$ and $P_j$ correct processes that both output $\babdel(\bl)$ and $\babdel(\bl')$.  $P_i$ delivers \bl before $\bl'$ if and only if $P_j$ delivers $\bl$ before $\bl'$.
  \item[External validity:] If a correct process outputs $\babdel(\bl)$, then $\VB(\bl)=\true$.
 \end{description}
\end{definition}

The block-based atomic broadcast abstraction can be implemented by protocols based on different approaches. These difference are not captured in Definition~\ref{def:bab}, but can relevant for the performance of the protocol. The two families of protocols of interest for this paper are \emph{chain protocol} and \emph{DAG protocol} protocols. The distinguishing factor between them lies in the set of references to previously mined blocks. Specifically, for a given block \bl, we denote the set of \babmine blocks referenced by \bl as $\op{parents}(\bl)$, commonly known as the \emph{parents} of \bl. Furthermore, the set of \babmine blocks reachable through references from \bl is represented as $\op{ancestors}(\bl)$ and is often referred to as the \emph{ancestors} of \bl. A block \bl is a \emph{descendant} of its ancestors. A block with no descendants is also called \emph{leaf}.

\begin{figure}
    \centering
    \includegraphics[width=0.9\textwidth]{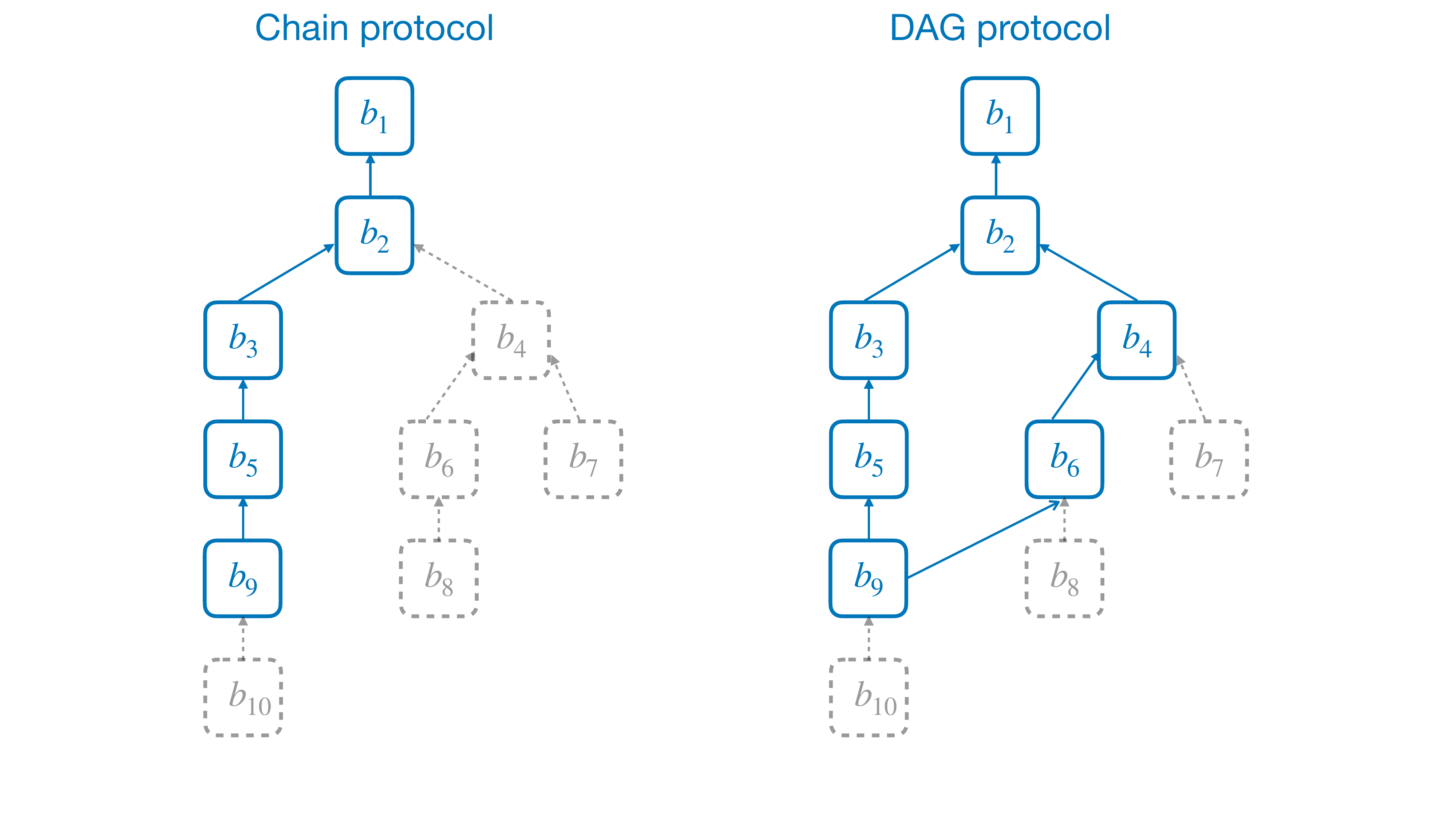}
    \vspace*{-4ex}
    \caption{Comparison between a chain protocol and a DAG protocol. Blocks in blue (continuous lines)
    are the \op{bab-delivered} blocks, whereas grey (dashed) blocks are \op{bab-mined} but not \op{bab-delivered}. The protocol on the left is a chain protocol, each block refers to exactly one block and there is a block ($\bl_9$) such that every currently \op{bab-delivered} block is $\bl_9$ or an ancestor of it. The protocol on the right is a DAG protocol, block $\bl_9$ references multiple blocks.}
   \label{fig:example}
  \end{figure}

\begin{definition}[Chain protocol, DAG protocol] 
  \label{def:chain protocol}
  A block-based atomic broadcast protocol $\Pi$ is a \emph{DAG protocol} protocol if $\Pi\op{-mined}$ blocks contain references to other $\Pi\op{-mined}$ blocks, meaning that the set of references is not empty. $\Pi$ is a \emph{chain protocol} protocol if every $\Pi\op{-mined}$ block refers to exactly one $\Pi\op{-mined}$ block and for every honest process $P_i$ there is a $\Pi\op{-delivered}$ block \bl such that every $\Pi\op{-delivered}$ by $P_i$ is $\bl$ or in $\op{ancestors}(\bl)$. In essence, $\Pi\op{-delivered}$ blocks form a chain.
  \end{definition}
  Figure~\ref{fig:example} illustrates an example of both chain and DAG protocols.

  To set the stage, we make the assumption that both chain and DAG protocols begin with an initial, hard-coded block referred to as the \emph{genesis} block. This genesis block is special in that it possesses an empty set of references. It is important to note that, according to Definition~\ref{def:chain protocol}, chain protocols inherently are DAG protocols. The blocks mined in chain protocols produce a \emph{tree}, a particular kind of DAG. Therefore, for the remainder of this paper, we will use the term ``DAG protocol'' to encompass both DAG protocol and chain protocols, acknowledging this inclusion. 

  One significant implication of abstracting DAG protocols as block-based atomic broadcast (Definition~\ref{def:bab}) is that the protocol must define a function that operates on the directed acyclic graph (\emph{DAG}) that produces a list of delivered blocks. It is worth mentioning that certain DAG protocols, such as the original Avalanche protocol~\cite{DBLP:journals/corr/abs-1906-08936,DBLP:conf/opodis/Amores-SesarCT22}, do not output an ordered list of transactions but the list output by different processes may differ up to permutation. While DAG protocols can also be modeled as generic broadcast~\cite{DBLP:conf/wdag/PedoneS99}, situations arise where complete transaction ordering, as seen in calls to smart contracts, becomes necessary. For the purposes of this paper, we focus on protocols that can be effectively modeled as block-based atomic broadcast. The results we derive in this context generalize straightforwardly to protocols modelled as generic broadcast.

  \section{Model}

  DAG protocols base their security on different techniques such as \emph{proof of work} (\emph{PoW}), \emph{proof of stake} (\emph{PoS})~\cite{DBLP:conf/eurocrypt/DavidGKR18}, \emph{proof of space-time} (\emph{PoST})~\cite{chia}, or \emph{proof of elapsed time} (\emph{PoET})~\cite{DBLP:conf/indocrypt/BowmanDMM21}. For the sake of simplicity, we restrict our model to PoW. Nevertheless, our model can readily be extended to incorporate other techniques.

  \paragraph{Processes.}
  Consistent with prior research, our protocol operates without explicit knowledge of the number or identities of the processes. The processes themselves remain unaware of these details as well. We assume a static network consisting of $n$ processes, where up to $f$ processes to be corrupted by the adversary, thereby exhibiting arbitrary behavior.

  \paragraph{Blocks.}
A transaction $\var{tx}$, comprises a set of \emph{inputs}, a set of \emph{outputs}, and a collection of digital signatures, as in Bitcoin~\cite{nakamoto2019bitcoin}. Transactions have size $|\var{tx}|$, and they are grouped into blocks, as introduced in Definition~\ref{def:bab}. Each block encompasses a specific number of transactions, denoted as $m$, a number of references to previously \op{bab-mined} blocks, quantified as $n_\var{refs}$, and  further parameters essential for the proper execution of protocol $\Pi$. It is noteworthy that the size of a reference, represented as $|\var{ref}|$, is significantly smaller than that of a transaction, for simplicity, we consider it to be negligible. We reiterate that protocol $\Pi$ defines external validity predicates, $\var{VT}()$ and $\var{VB}()$, responsible for determining the \emph{validity} of a transaction or block.

\paragraph{Network.}
A \emph{diffusion functionality} implements communication among the processes, which is structured into \emph{synchronous} rounds. The functionality keeps a distinct $\op{RECEIVE}_i$ string for each process $P_i$ and makes it available to $P_i$ at the start of every round. The purpose of the string $\op{RECEIVE}_i$ is to serve as a repository for all the messages received by $P_i$.

When a process, say $P_i$, instructs the diffusion functionality to \emph{broadcast} a set of message, it signifies that $P_i$ has ``completed its round''. In response, the functionality marks $P_i$ as having completed its operations for that specific round. The adversary, whose actions are described in detail below, possesses the ability to access the string of any process at any point during the execution. Additionally, the adversary can observe every message broadcast by any process instantaneously. Furthermore, the adversary has the capability to insert messages directly and selectively into $\op{RECEIVE}_i$ for any process $P_i$, ensuring that only $P_i$ receives the message at the outset of the following round. This behavior models what is often termed a \emph{rushing} adversary.

Once all non-corrupted processes have concluded their respective rounds, the diffusion functionality aggregates all messages that were broadcast by non-corrupted processes during that round. These aggregated messages are then appended to the $\op{RECEIVE}_i$ strings for all processes, this is the reason of the name \emph{synchronous} rounds. Subsequently, each non-corrupted process updates its local view at the conclusion of every round. If a non-corrupted process $\Pi\op{-mines}$ a block in round~$r$, all processes receive the $\Pi\op{-mined}$ block by the subsequent round $r+1$.

Furthermore, even if the adversary causes a block to be received selectively by only some non-corrupted processes in round $r$, the block is received by all non-corrupted processes by round $r+2$. The update of the local view also encompasses the $\Pi\op{-delivery}$ of blocks that meet a given criteria define by protocol $\Pi$.

\paragraph{Adversary.} The adversary can corrupt up to $f$ processes at the beginning of the execution. These corrupted processes may deviate arbitrarily from  the protocol, adhering to the instructions from the adversary. Additionally, the adversary wields control over the \emph{diffusion functionality}. The adversary can schedule the delivery of messages, read the contents of the $\op{RECEIVE}_i$ string for every process at any point during the execution, and directly write messages into the $\op{RECEIVE}_i$ of any process. The adversary signals the conclusion of her round by transmitting a specially designated message.

\paragraph{Round structure.} 
At the beginning of the round, process $P_i$ reads the messages in its input string $\op{RECEIVE}_i$. Then, $P_i$ proceeds to update its internal state in accordance with the received messages and performs a set of actions defined by protocol $\Pi$. Such actions include the $\Pi\op{-delivery}$ of blocks. $P_i$ concludes the round by broadcasting a set of messages to the other processes.

\subsection{Abandoned blocks}
\label{sec:abandoned}
\begin{definition}
  An \emph{execution} is a history with an entry for each round containing the actions, a list of received messages, and a list of sent messages by each process in that round.
 \end{definition} 
 
 While an event may be theoretically possible within an execution, its occurrence might have a probability of zero. For instance, consider an algorithm that continuously flips an unbiased coin indefinitely. There could be an execution where all outcomes are heads, but the probability of this specific sequence of events happening is zero, as it is the limit of an infinite execution. 
 
 To circumvent these issues, we introduce the concept of a \emph{partial execution}.
 
  \begin{definition}
   Given a protocol $\Pi$, the set of \emph{$\lambda$-partial executions} $\Phi_\lambda$ is defined to be the set of $\lambda$-prefixes of all executions of protocol $\Pi$. A \emph{partial execution} is an execution that belongs to $\Phi_\lambda$ for some $\lambda\in\mathbb{N}$.
  \end{definition}
 
 \begin{definition}
   Given an execution \CE of a block-based atomic broadcast protocol $\Pi$, an \emph{abandoned} block in \CE is is an honestly $\op{bab-mined}$ block $b$ such that \bl is not $\op{bab-delivered}$ in \CE. 
 \end{definition}

 It is important to note that the validity property defined in block-based atomic broadcast (Definition~\ref{def:bab}) does not guarantee that every $\op{bab-mined}$ block will eventually be $\op{bab-delivered}$. Instead, this property ensures that for each $\op{bab-broadcast}$ transaction, there exists at least one $\op{bab-delivered}$ block that contains it. The concept of abandoned blocks is a significant concern in the context of such protocols. Abandoned blocks have been honestly $\op{bab-mined}$ but are never $\op{bab-delivered}$. The existence of abandoned blocks can severely impact the performance of a chain protocol or DAG protocol.

 \begin{definition}
 A protocol $\Pi$ \emph{permits abandoned blocks} if there exist a block $b$ and a partial execution \CE such that: $b$ is abandoned in any extension of \CE.  
 \end{definition}
 
 \begin{remark}
   \label{remark:probabilities}
   Note that given a protocol that permits abandoned blocks, the probability, taken over the randomness of the protocol, of having at least one abandoned block in an execution is greater than zero, since partial executions happen with non-zero probability.
 \end{remark}

 Determining whether a given protocol $\Pi$ permits abandoned blocks or not can be a challenging task and, in some cases, may not be computable due to the need to simulate potentially infinitely long executions. However, for certain protocols like Bitcoin~\cite{nakamoto2019bitcoin}, the existence of abandoned blocks is a direct consequence forks occurring among honest miners. This phenomenon is formalized in the following definition.

 \begin{definition}
   Given an execution $\CE$ of a given protocol $\Pi$, a round $r$ \emph{forked} if protocol $\Pi$ outputs two events $\babmine(b,P_i)$ and $\babmine(b',P_j)$ in round $r$ at two distinct honest processes $P_i$ and $P_j$. A protocol with a forked round in at least one partial execution is a \emph{forkable protocol}.
 \end{definition}

 \begin{lemma}
  \label{lemma:forkable}
   A forkable chain protocol $\Pi$ permits abandoned blocks.
 \end{lemma}
 \begin{proof}
   Given a forkable protocol $\Pi$, there  exist a round $r$ in which two different honest processes output  events $\babmine(b,P_i)$ and $\babmine(b',P_j)$. In particular $b\neq b'$ because their miners are different. $\Pi$ is also a chain protocol. thus both $b$ and $b'$ have a unique reference to previously \babmine blocks, so they cannot reference each other. Another implication of $\Pi$ being a chain protocol is that at any point in the execution in the protocol there exists a \babmine block $b^*$ such that every \op{bab-delivered} is in $\op{ancestors}(b^*)$. Since every block only contains a single reference and $b$ and $b'$ do not refer each other, we conclude that no honest processes can \op{bab-deliver} both $b$ and $b'$ simultaneously.
 \end{proof}

 Transactions that were originally included in abandoned blocks must be re-included in subsequent blocks to maintain the validity property (Definition~\ref{def:bab}). This re-inclusion consumes space in new blocks and has implications for both latency and throughput, as we formalize below.

\subsection{Throughput and latency}
\label{sec:latency}

\begin{definition}
Given a block-based atomic broadcast protocol $\Pi$, an adversary \CA, and an execution $\CE$, we define the \emph{throughput} of $\Pi$ in the presence of \CA in execution $\CE$ as the average number of $\op{bab-delivered}$ blocks per round and we denote by $\op{throughput}(\Pi,\CA,\CE)$. 
\end{definition}

\begin{definition}
Given a block-based atomic broadcast protocol $\Pi$, the \emph{throughput of $\Pi$} is defined to be $\displaystyle\op{throughput}(\Pi):=\inf_{\CA} \E[\op{throughput}(\Pi,\CA,\CE)]$, i.e., 
the infimum over all the possible adversaries \CA of the average over the randomness $\Pi$ of $\op{throughput}(\Pi,\CA,\CE)$ over all the possible executions.  
\end{definition}

\begin{definition}
The \emph{goodput} of protocol $\Pi$ is defined to be throughput of $\Pi$ in the presence of an adversary that follows the instructions of the protocol. 
\end{definition}

\begin{definition}
Given a block-based atomic broadcast protocol $\Pi$, an adversary \CA, an execution $\CE$, and a transaction \tx, we define \emph{latency} of \tx in the presence of adversary \CA in execution $\CE$ as the number of rounds since $\var{tx}$ is $\op{bab-broadcast}$ until the first block containing $\var{tx}$ is $\op{bab-delivered}$, and we denote it by $\op{latency}(\Pi,\CA,\CE,\tx)$. We define the \emph{latency of $\Pi$} to be the average number of rounds, over the transactions $\var{tx}$ in execution $\CE$, since  $\var{tx}$ is $\op{bab-broadcast}$ until the first block containing $\var{tx}$ is $\op{bab-delivered}$ and denote it by $\op{latency}(\Pi,\CA,\CE)$.
\end{definition}

\begin{definition}
  Given a block-based atomic broadcast protocol $\Pi$, The \emph{latency} of protocol $\Pi$ is defined as $\displaystyle\op{latency}(\Pi)=\sup_{\CA} \E[\op{latency}(\Pi,\CA,\CE)]$, i.e., the supremum over all the possible adversaries \CA of the average over the randomness of the protocol of the $\op{latency}(\Pi,\CA,\CE)$ over the possible executions $\CE$. 
\end{definition}

\section{The throughput closure}

We introduce a novel construction designed to enhance a given DAG protocol~$\Pi$. This construction results in a DAG protocol, which we call \emph{the throughput closure of $\Pi$} and denote by $\Pi^\prime$. Protocol $\Pi'$ possesses the unique property of ensuring that every honestly $\op{bab-mined}$ block is eventually $\op{bab-delivered}$. The mechanism by which protocol $\Pi^\prime$ accomplishes this feat involves the incorporation of additional references to blocks. For any given block \bl, protocol $\Pi^\prime$ defines the set $\op{abandoned}(\bl)$ as the collection of valid blocks that will not be $\Pi\op{-delivered}$ if \bl is to be $\Pi\op{-delivered}$. The block mining and delivery routines of the throughput closure $\Pi^\prime$  are built on top of their counterparts in $\Pi$.

\paragraph{Overview.}
As shown in Algorithm~\ref{algo:construction}, when an honest process $P_i$ $\Pi\op{-mines}$ a block $\bl$, process $P_i$ also $\Pi'\op{-mines}$ the same block. However, in $\Pi^\prime$, the block $\bl$ includes an additional set of references to the blocks in the set $\op{abandoned}(\bl)$. 

The modified delivery routine operates as follows: when a block \bl would be $\Pi\op{-delivered}$, all valid blocks in the set $\op{abandoned}(\bl)$  are $\Pi'\op{-delivered}$ in a fixed topological order immediately before \bl. This topological sort allows to order non $\Pi\op{-delivered}$ blocks with respect to $\Pi\op{-delivered}$ blocks deterministically according to the references included in the $\Pi\op{-delivered}$ blocks. This is a crucial aspect as establishing a total order in a DAG can be generally challenging due to different processes having different partial views of the DAG. The topological sort $\tau$ ensure that all processes that have received block $\bl$ agree on the same order.
A canonical example for \emph{topological sort $\tau$} is to order the blocks in $\op{abandoned}(\bl)$ according to their \emph{depth} in the DAG, distance to genesis, breaking the ties according to the hash of the block. Note that if an adversary creates a block with low depth, it will be only $\Pi\op{-delivered}$ when deeper block references it, thus the adversarial block is $\Pi'\op{-delivered}$ concurrently with deeper blocks.

\begin{figure}
  \centering
  \includegraphics[width=0.9\textwidth]{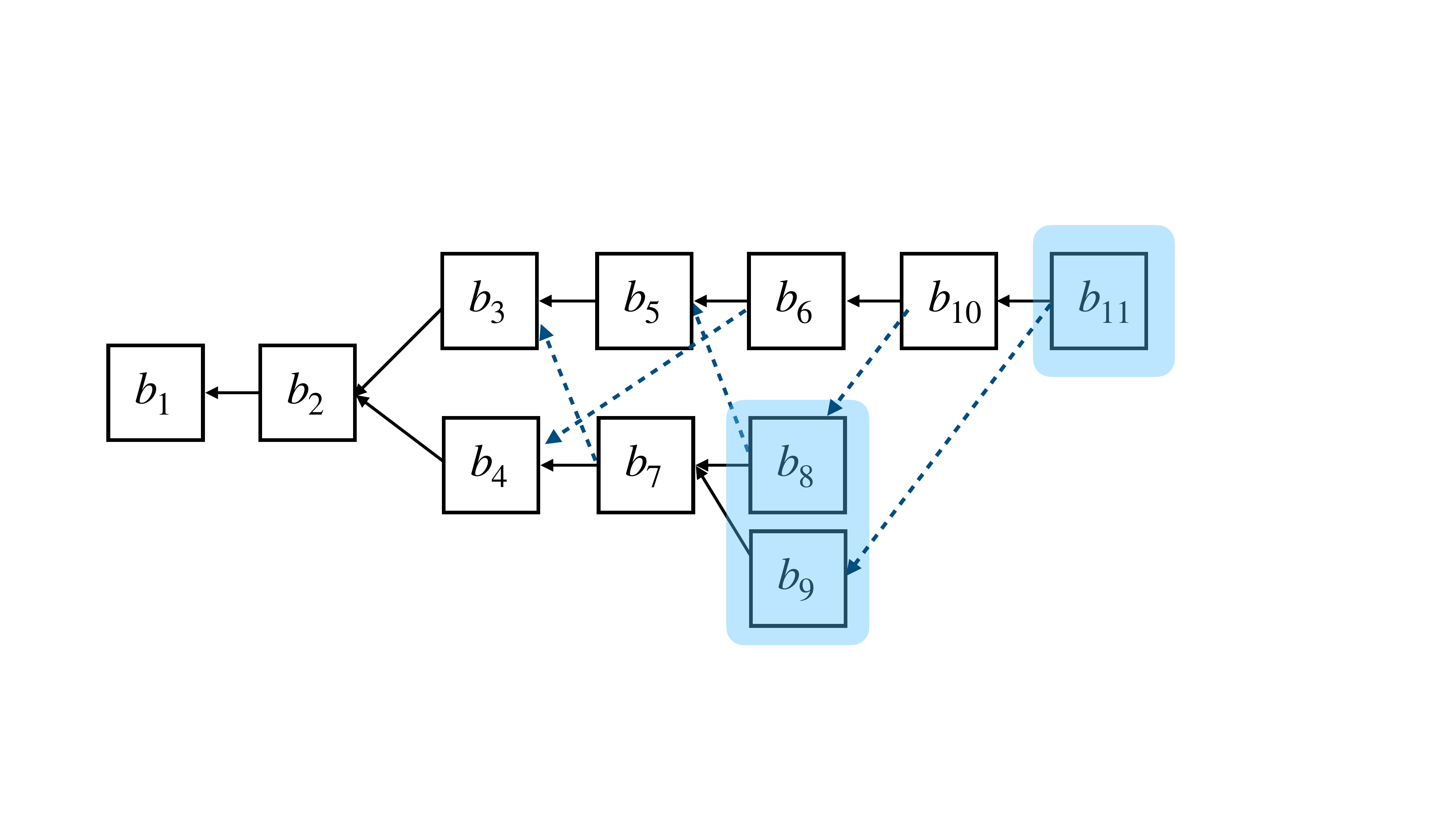}
  \vspace*{-4ex}
  \caption{An example of our construction applied to Nakamoto consensus. The full lines denote the references of the Nakamoto consensus and the blue dashed lines denote the extra references included by the throughput closure.
  According to Nakamoto consensus, the main chain is the chain $b_1\cdots b_{11}$ and blocks $b_4,b_7,b_8$, and $b_9$ are abandoned. Looking at $b_{11}$, the set $\op{abandoned}(b_{11})$ is formed by block $b_9$. Blocks $b_4,b_7$, and $b_8$ are not part of the $\op{abandoned}(b_{11})$ because $b_{10}$ already references them. When delivering $b_{11}$, block $b_9$ would be delivered between $b_{10}$ and $b_{11}$.}
 \label{fig:example2}
\end{figure}

Constructing the set $\op{abandoned}(\bl)$, even when it can be computed, may be challenging task, as we explained above. However, given a chain protocol $\Pi$ the set $\op{abandoned}(\bl)$ becomes trivial to compute as it is formed by every block that is not an ancestor of $\bl$. Furthermore, the set of references to $\op{abandoned}(\bl)$ are the leaves of the DAG, with the exception of \bl. As an illustrative example, Figure~\ref{fig:example2} shows the application of this construction within the context of Bitcoin. If we consider $\Pi$ to be GHOST protocol~\cite{DBLP:conf/fc/SompolinskyZ15}, we recreate the Conflux protocol~\cite{DBLP:conf/usenix/LiLZYWYXLY20}. Including references to the leaves in the DAG is the precise method for referring to the set $\op{abandoned}(\bl)$ with a chain protocol~$\Pi$. The same approach can be used with DAG protocols. This approach may be computationally cheaper than than computing the leaves in set $\op{abandoned}(\bl)$, however, some blocks may be referenced when there is no need, adding redundancy of references. Further insights into this alternative approach are provided below.

\label{sec:detailed}

  \begin{algo*}
    \vbox{
    \small
    \begin{numbertabbing}
      xxxx\=xxxx\=xxxx\=xxxx\=xxxx\=xxxx\=MMMMMMMMMMMMMMMMMMM\=\kill
      \textbf{Implements:} block-based atomic broadcast $\Pi'$ \\
      \textbf{Uses:} block-based atomic broadcast $\Pi$ \\
      \> \hspace{1.2 mm}topological sort $\tau$\\
      \\

      \textbf{State:}\\
      \> $\CD' \gets \emptyset$ \label{}\\
      \> $b_\ell'\gets [\ ]$\label{}\\
      \\
      \textbf{upon event} $ \Pi'\op{-broadcast}(\tx)$ \textbf{do}\label{line:bcast-begin}\\
      \>\textbf{invoke} $\Pi\op{-broadcast}(\tx)$ \label{line:bcast-end}\\
      \\

      \textbf{upon event} $\Pi\op{-mined}(\bl, P_j) $ \textbf{do}\label{line:mined-begin}\\
      \>\textbf{if} $P_i=P_j$ \textbf{then}\label{}\\
      \>\>$\var{weak}\gets \op{leaves}(\op{abandoned}(\bl,\CD'))$\label{line:extra1}\\
      \>\>$\bl'\gets\bl$\label{}\\
      \>\>$bl'.\op{wrefs}\gets \var{weak}$ \label{line:extra2}\\
      \>\> $\CD'\gets \CD' \cup \{\bl'\}$\label{line:setmined}\\
      \>\>\textbf{invoke} $ \Pi'\op{-mined} (\bl', P_i) $  \label{line:mined-end}\\ 
      \\

      \textbf{upon event} $\Pi'\op{-mined}(\bl', P_j)$  \textbf{do}\label{line:receive-begin}\\
      \> \textbf{if}  $\VB'{}(\bl')$ \textbf{then}\label{}\\
      \>\>$\CD'\gets \CD' \cup \{\bl'\}$\label{line:receive-end}\\
      \\

      \textbf{upon event} $ \Pi\op{-deliver}(\bl)$ \textbf{do}\label{line:babdel-begin}\\
      \>$\var{ready}\gets \op{ancestors'}(\bl')\setminus \op{ancestors'}(\bl_\ell')$\label{line:setdif}\\
      \>$\bl_\ell'\gets \bl'$\label{line:update}\\
      \>\textbf{for} $\bl^* \in \tau(\var{ready})$  \textbf{do}\label{line:topological}\\
      \>\> \textbf{invoke} $\Pi'\op{-deliver}(\bl^*)$\label{line:deliver-end}\\
      \\

      \textbf{function} $\op{abandoned}(\bl,\CD')$ \textbf{:}\label{l:abandoned}\\
      \>\textbf{return} $\{\bl'\in\CD':\bl'\not\in \op{ancestors'} (\bl)\land \op{incompatible}(\bl,\bl') \}$\label{}\\
      \\
      \textbf{function} $\VB'{}(\bl')$ \textbf{:} \label{l:external1}\\
      \> \textbf{return} $\VB{}(\bl)\land \exists\ \var{tx}\in \bl':\op{undelivered}(\var{tx})$ \label{l:external2}
    \end{numbertabbing}
    }
    \caption{Protocol $\Pi'$ for process~$P_i$.}
    \label{algo:construction}
  \end{algo*}

  \begin{algo*}
    \vbox{
    \small
    \begin{numbertabbing}
      xxxx\=xxxx\=xxxx\=xxxx\=xxxx\=xxxx\=MMMMMMMMMMMMMMMMMMM\=\kill

  \textbf{upon event} $\Pi\op{-mined}( \bl, P_j)$ \textbf{do}\label{line:babmine-begin}\`// Greedy approach\\
  \>\textbf{if} $P_i=P_j$ \textbf{then}\label{}\\
  \>\>$\bl'\gets\bl$\label{}\\
  \>\>$\var{weak}\gets \op{leaves}(\{\bl'\in\CD': \bl'\not \in \op{ancestors}(\bl')\})$\label{}\\
  \>\>$bl'.\op{refs}\gets bl'.\op{refs}|| \var{weak}$ \label{}\\
  \>\> $\CD'\gets \CD' \cup \{\bl'\}$\label{}\\
  \>\textbf{invoke} $ \Pi'\op{-mined}(\bl', P_i)$  \label{line:loop-through-mined-end}
\end{numbertabbing}
}
\caption{Greedy approach for process $P_i$.}
\label{algo:greedy}
\end{algo*} 

\paragraph{Detailed description.}
We describe the execution of the protocol from the perspective of an honest process $P_i$.
When honest process $P_i$ $\Pi'\op{-broadcasts}$ a transaction $\var{tx}$, it invokes $\Pi\op{-broadcast}(\tx)$ (L\ref{line:bcast-begin}--\ref{line:bcast-end}). Notably, the broadcast of transactions occurs exactly as it does in protocol $\Pi$.
When $P_i$ triggers event $ \Pi\op{-mined} (\bl,P_i)$ (L\ref{line:mined-begin}--\ref{line:mined-end}), it initially computes the set $\op{abandoned}(\bl)$ locally. To $\Pi'\op{-mine}$ a new block $\bl'$, $P_i$ augments \bl by adding extra references to the leaves of the set $\op{abandoned}(\bl)$ (L\ref{line:extra1}--\ref{line:extra2}). Subsequently, $P_i$ adds $\bl'$ to the set of mined blocks $\CD'$ (L\ref{line:setmined}) and triggers the event $ \Pi'\op{-mined}( \bl', P_i)$ (L\ref{line:mined-end}).

When event $ \Pi'\op{-mined}( \bl', P_j)$ is triggered, $P_i$ verifies the $\Pi'$-validity of $\bl'$ and incorporates it into its local view (L\ref{line:receive-begin}--\ref{line:receive-end}). So far, the execution of $\Pi'$ closely parallels that of $\Pi$. However, the key distinction lies in the delivery of blocks (L\ref{line:babdel-begin}--\ref{line:deliver-end}). When event $ \Pi\op{-deliver}(\bl)$ occurs, $P_i$ searches for the block $\bl'$ associated with \bl. $P_i$ then assembles the set $\var{ready}$, which comprises the blocks to be $\Pi'\op{-delivered}$ (L\ref{line:setdif}). This set is computed as the set-difference between the ancestors of block $\bl'$ and the ancestors of the last delivered block $\bl_l'$. $P_i$ subsequently updates the last delivered block to be $\bl'$ (L\ref{line:update}). Finally, $P_i$ applies a topological sorting algorithm $\tau$ to the set $\var{ready}$ and $\Pi'$-delivers them accordingly (L\ref{line:topological}--\ref{line:deliver-end}).

A block $\bl'$ is deemed valid (L\ref{l:external1}--\ref{l:external2}) within protocol $\Pi'$ if it satisfies two conditions: firstly, its associated block \bl must be $\Pi\op{-valid}$, and secondly, it must contain at least one $\Pi'$-valid transaction.

 Algorithm~\ref{algo:greedy} presents a greedy version of $\op{abandoned}(\bl)$. In this approach, a process $P_i$ adds references to $\bl'$ for every block that is not already an ancestor of $\bl$ within protocol $\Pi$.

\paragraph{}The throughput closure mirrors protocol $\Pi$ when the set $\op{abandoned}(\bl)$ is empty for every block, indicating that the protocol does not permit the existence of abandoned blocks. However, if $\Pi$ permits abandoned blocks, then there exists some executions of $\Pi$ with a block \bl such that $\op{abandoned}(\bl)\neq\emptyset$, and the throughput closure diverges from the original protocol. The implementation of the throughput closure does entail an increase in local computation for processes. Specifically, processes need to scan the DAG and append a set of references to all leaves in $\op{abandoned}(\bl)$ to the currently mined block \bl. The computational complexity of determining $\op{abandoned}(\bl)$ can vary depending on the protocol, as discussed earlier. However, in the case of chain protocols, this set is relatively straightforward to compute. A process simply adds references to every leaf of a chain that has not been referenced by an ancestor.

\section{Analysis}
\subsection{Security analysis}
\label{sec:security}
\begin{theorem}
    \label{theo:construction}
    Given protocol DAG protocol $\Pi$ implementing block-based atomic broadcast, its throughput closure $\Pi^\prime$ also implements block-based atomic broadcast.
\end{theorem}

\begin{proof}
    
  We demonstrate that the throughput closure $\Pi^\prime$ implements block-based atomic broadcast by leveraging the fact that $\Pi$ does. Throughout this proof, we assume the perspective of an honest process~$P_i$.
  \begin{description}

  \item[Validity:] Assume that an honest process $P_j$ $\Pi'\op{-broadcasts}$ a given transaction $\var{tx}$. By construction, process $P_j$ does so by invoking $\Pi\op{-broadcast}$ transaction $\var{tx} $ (L\ref{line:bcast-begin}--\ref{line:bcast-end}). The validity property of protocol $\Pi$ guarantees that process $P_i$ eventually $\Pi\op{-delivers}$ a block \bl containing transaction $\var{tx}$. Process $P_i$, by definition of the protocol, $\Pi'\op{-delivers}$ the block $\bl'$ consisting of block \bl with the addition of the extra set of references (L\ref{line:babdel-begin}--\ref{line:deliver-end}). If every transaction contained in $\bl'$ is invalid, the block is not $\Pi'\op{-deliver}$. In the case of block $\bl'$, the validity check can only fail if transaction $\var{tx}$ fails the validity predicate. Since $\var{tx}$ is $\Pi'\op{-broadcast}$, the external validity predicate is satisfied unless some block containing $\var{tx}$ has been $\Pi'\op{delivered}$.

  We conclude that for any honestly $\Pi'\op{-broadcast}$ transaction \var{tx}, $P_i$ eventually $\Pi'\op{-delivers}$ a block $\bl'$ containing $\var{tx}$, thus validity property of protocol $\Pi'$ is satisfied.

  \item [Integrity:] Process $P_i$ only $\Pi'\op{-delivers}$ blocks that it $\Pi'\op{-delivers}$ or ancestors of those contained in the set \CD (L\ref{line:babdel-begin}--\ref{line:deliver-end}). A block $\bl'$ enters the set \CD only after an invokation of $ \Pi\op{-mined}( \bl, P_j)$. We conclude that every $\Pi'\op{-delivers}$ has previously been $\Pi'\op{-mined}$.

  \item [Agreement:] Consider a block $\bl'$ that is $\Pi'\op{-delivered}$ by process $P_i$. We consider two different cases: when \bl whether \bl is $\Pi\op{-delivered}$ or not. On the one hand, if \bl is $\Pi\op{-delivered}$ by process $P_i$, every honest process eventually $\Pi\op{-delivers}$ \bl, thus $\Pi'\op{-delivers}$ $\bl'$ as a consequence (L\ref{line:babdel-begin}--\ref{line:deliver-end}). On the other hand, if block $bl'$ is $\Pi'\op{-delivered}$ as a consequence of another block $\bl^*$ is $\Pi'\op{-delivered}$. The same reasoning as above applies to block $\bl^*$, which implies the eventual $\Pi'\op{-delivery}$ of block $\bl'$.

  \item[Total order:] Consider two $\Pi'\op{mined}$ blocks $\bl_1'$ and $\bl_2'$ and two honest processes $P_i$ and $P_j$ that $\Pi'-\op{deliver}$ both blocks. We distinguish four cases based on whether blocks $\bl_1$ and $\bl_2$ are $\Pi\op{-delivered}$ or not.

  Assume that both $\bl_1$ and $\bl_2$ are $\Pi\op{-delivered}$. Note that in the view of any honest process the order in which blocks $\bl_1$ and $\bl_2$ are $\Pi\op{-delivered}$ is the same as blocks $\bl_1'$ and $\bl_2'$ are $\Pi'\op{-delivered}$ (L\ref{line:babdel-begin}--\ref{line:deliver-end}). Due to the total order property of protocol $\Pi$, process $P_i$ $\Pi\op{-delivers}$ block $\bl_1$ and $\bl_2$ in the same order as process $P_j$, thus both processes $\Pi'\op{-deliver}$ blocks $\bl_1$ and $\bl_2$.

  If either $\bl_1'$ or $\bl_2'$ are $\Pi'\op{-delivered}$ as a consequence of another block $\bl_3$ being $\Pi\op{-delivered}$. Since the set of blocks that are $\Pi'\op{-delivered}$ as consequence of block $\bl_3'$ are $\Pi'\op{-delivered}$ immediately before $\bl_3'$, any block $\bl'$ $\Pi'\op{-delivered}$ before (after) $\bl'$ is also $\Pi'\op{-delivered}$  before (after) the set of blocks $\Pi'\op{-delivered}$ as a consequence of $\bl'$. The same reasoning as above applies to this case. We conclude that $P_i$ also $\Pi'\op{delivers}$ both $\bl_1'$ or $\bl_2'$ in the same order as $P_j$.

  The only case left is when both $\bl_1'$ and $\bl_2'$ are  $\Pi'\op{-delivered}$ as a consequence of two blocks $\bl_3'$ and $\bl_4'$ being  $\Pi'\op{-delivered}$. If $\bl_3'$ and $\bl_4'$ are different the case is the same as before. If $\bl_3'$ and $\bl_4'$, both $P_i$ and $P_j$ use the topological order to determine in which order to $\Pi'\op{-delivered}$. Since the topological sorting is deterministic and depends only on block $\bl_3'$, both $P_i$ and $P_j$ $\Pi'\op{-deliver}$ $\bl_1'$ and $\bl_2'$ in the same order.

   \item[External validity:] The external validity property is imposed by lines L\ref{l:external1}--\ref{l:external2}.
 
  \end{description}
      
\end{proof}

\subsection{Throughput and latency}
Theorem~\ref{theo:construction} states that the throughput closure $\Pi'$ maintains the safety and liveness properties the original protocol $\Pi$. In this section, we delve into a comparative analysis of the performance aspects, through throughput and latency, between $\Pi'$ and $\Pi$. 
It is important to note that both throughput and latency definitions take into account adversarial behavior, and the connection between the adversarial behavior of $\Pi'$ and $\Pi$ is discussed in the following remark.

\begin{remark}
\label{remark:adversaries}
Note that given an adversary $\CA'$ for protocol $\Pi'$,  an adversary $\CA$ for protocol $\Pi$ can be constructed by merely removing the extra references from any block that $\CA'$  $\Pi'\op{-mines}$. Additionally, given an adversary $\CA$ for protocol $\Pi$, it can also be regarded as an adversary for protocol $\Pi'$, as every action taken by $\CA$ in protocol $\Pi$ is allowed in protocol $\Pi'$.
\end{remark}

\begin{definition}
\label{def:equivalent}
Given an execution $\CE'$ and an adversary $\CA'$ for protocol $\Pi'$, we define the equivalent execution of protocol $\Pi$ as the execution $\CE'$ without the extra references in each block and adversary $\CA$  as discussed in Remark~\ref{remark:adversaries}.
\end{definition}

\begin{lemma}
\label{lemma:latency}
Given a DAG protocol $\Pi$, its throughput closure $\Pi^\prime$ achieves the same or lower latency as $\Pi$. 
\end{lemma}
\begin{proof}
Consider an execution $\CE'$, an adversary $\CA'$ for protocol $\Pi'$, and a transaction $\var{tx}$ that has not already been $\Pi'\op{-delivered}$. Denote by $\CE$ the equivalent execution (Definition~\ref{def:equivalent}) of protocol $\Pi$.
Note that by definition of $\Pi'$, \tx has not been $\Pi\op{-delivered}$ either (L\ref{line:babdel-begin}).
Protocol $\Pi'$ has two different mechanisms to $\Pi'\op{-deliver}(\tx)$. 

On the one hand, if an event $\Pi\op{-deliver}(\bl)$ for a block \bl containing transaction \tx is triggered, then \bl is $\Pi'\op{-delivered}$ (L\ref{line:babdel-begin}). In this case, $\op{latency}(\Pi',\CA',\CE',\tx)$ is the same as $\op{latency}(\Pi,\CA,\CE,\tx)$.

On the other hand, if an event $\Pi\op{-deliver}(\bl')$ for a block $\bl'$ that does not contains \tx but is descendent of a block $\bl$ containing \tx., then block $\bl'$ is $\Pi'\op{-delivered}$ immediately before \bl (L\ref{line:setdif}). In this case, $\op{latency}(\Pi',\CA',\CE',\tx)$ is strictly smaller than $\op{latency}(\Pi,\CA,\CE,\tx)$.

Both cases discussed above imply that for every adversary, execution, and transaction, the latency of both protocols satisfy $\op{latency}(\Pi',\CA',\CE',\tx)\leq\op{latency}(\Pi,\CA,\CE,\tx)$. Hence, $\op{latency}(\Pi')\leq \op{latency}(\Pi)$
\end{proof}

The next result clarifies the motivation for the term throughput closure.
\begin{lemma}
\label{lemma:throughput}
Given a DAG protocol $\Pi$, then $\op{throughput}(\Pi')\geq \op{throughput}(\Pi)$, and if $\Pi$ permits abandoned blocks, then $\op{throughput}(\Pi')> \op{throughput}(\Pi)$.
\end{lemma}
  
\begin{proof}
Consider an execution $\CE'$, an adversary $\CA'$ for protocol $\Pi'$, and a transaction $\var{tx}$ that has not already been $\Pi'\op{-delivered}$. Denote by $\CE$ the equivalent execution (Definition~\ref{def:equivalent}) of protocol $\Pi$.

On the one hand, if there is no abandoned block in the execution $\CE'$, then, the set $\op{abandoned}(\bl')$ is empty for every block $\bl'$. Thus, no extra reference is added at any point in the execution of $\Pi^\prime$ the executions $\CE$ and $\CE'$ identical. We conclude that $\op{throughput}(\Pi,\CA',\CE)=\op{throughput}(\Pi',\CA,\CE)$.

On the other hand, if there exists at least one abandoned block $\bl'$ in execution $\CE'$, then, the set $\op{abandoned}(\bl^*)$ is not empty for some block $\bl^*$ that is eventually $\Pi'\op{-delivered}$. When $\bl^*$ is $\Pi'\op{delivered}$ so is $\bl'$ (L\ref{line:setdif}). 
 
We conclude that $\op{throughput}(\Pi',\CA',\CE')\geq\op{throughput}(\Pi',\CA,\CE)$ for every possible adversary $\CA'$ and execution $\CE'$, thus $\op{throughput}(\Pi')\geq\op{throughput}(\Pi')$. Furthermore, if $\Pi$ permits abandoned blocks, there exists an $\lambda$-partial execution with a block $\bl$ that is abandoned in all its extensions. This means that the probability, over the randomness of the protocol, of having an abandoned block is strictly greater than zero (Remark~\ref{remark:probabilities}). Thus, $\E[\op{throughput}(\Pi',\CA',\CE')]>\E[\op{throughput}(\Pi,\CA,\CE)]$ for at least some adversary $\CA'$. We conclude by noticing that if an adversary $\CA^*$ prevents the exclusion of abandoned blocks, then $\op{throughput}(\Pi',\CA^*,\CE')>\op{throughput}(\Pi',\CA',\CE')$. Hence, we conclude that 

\begin{equation*}
  \begin{split}
    \displaystyle\op{throughput}(\Pi') & =\inf_{\CA'} \E[\op{throughput}(\Pi',\CA',\CE')]\\
    &>\inf_{\CA} \E[\op{throughput}(\Pi,\CA,\CE)]=\op{throughput}(\Pi).
  \end{split}
\end{equation*}

\end{proof}

\begin{corollary}
  \label{coro:goodput}
  Given a DAG protocol $\Pi$, then $\op{goodput}(\Pi')\geq \op{goodput}(\Pi)$. Furthermore, if $\Pi$ allows for the existence of abandoned blocks, then $\op{goodput}(\Pi')> \op{goodput}(\Pi)$.
\end{corollary}

\begin{proof}
Consider the proof of Lemma~\ref{lemma:throughput} limited to adversaries that follow the instructions of the protocol.
\end{proof}
  
  Note that every chain protocol trivially permits abandoned block. We can finally conclude that DAG protocols are strictly better then chain protocols.
  
  \begin{theorem}
    Given a chain protocol $\Pi$, there exists a DAG protocol $\Pi'$ such that: $\op{latency}(\Pi')\leq \op{latency}(\Pi)$ and $\op{throughput}(\Pi')>\op{throughput}(\Pi)$.
  \end{theorem}
  
  \begin{proof}
    Lemma~\ref{lemma:forkable} states that a chain protocol $\Pi$ permits abandoned blocks. Theorem~\ref{theo:construction} demonstrates that its throughput closure $\Pi'$ implements block-based atomic broadcast. Lemma~\ref{lemma:throughput} establishes that $\op{throughput}(\Pi')> \op{throughput}(\Pi)$. Finally, Lemma~\ref{lemma:latency} shows that $\op{latency}(\Pi')\leq \op{latency}(\Pi)$.
  \end{proof}
  
  \section*{Acknowledgments}

  This work has been funded by the Swiss National Science Foundation (SNSF)
  under grant agreement Nr\@.~200021\_188443 (Advanced Consensus Protocols).

\bibliography{references, dblpbibtex}
\bibliographystyle{ieeesort}

\end{document}